\documentclass[reprint,showpacs,amsfonts,amsmath,amssymb,prb]{revtex4-1}
\usepackage{graphicx}
\usepackage{dcolumn}
\usepackage{bm}

\begin{document}

\title{Spin fluctuations and pairing symmetry in A$_{x}$Fe$_{2-y}$Se$_{2}$: dual effect of the itinerant and the localized nature of electrons}

\author{Shun-Li Yu}
\author{Jia Guo}
\author{Jian-Xin Li}
\affiliation{National Laboratory of Solid State Microstructures and Department of Physics, Nanjing University, Nanjing 210093, China}

\date{\today}

\begin{abstract}
We investigate the spin fluctuations and the pairing symmetry in A$_{x}$Fe$_{2-y}$Se$_{2}$ by the fluctuation exchange approximation. Besides the on-site interactions, the next-nearest-neighbor antiferromagnetic coupling $J_{2}$ is also included. We find that both the itinerant and the localized natures of electrons are important to describe the recent experimental results of the spin fluctuations and the pairing symmetry. In particular,
 a small $J_{2}$ coupling can change the pairing gap from the $d$-wave symmetry to the $s$-wave symmetry. We have also studied the real-space structures of the gap functions for different orbits in order to gain more insight on the nature of the pairing mechanism.
\end{abstract}


\maketitle

\section{INTRODUCTION}

The discovery of a new family of iron-based superconductors (FeSC) A$_{x}$Fe$_{2-y}$Se$_{2}$ (A=K, Rb, Cs) with $T_c$ more than 30 K has attracted much attention \cite{RPB.82.180520,PRB.83.060512,JPCM.23.052203,EPL.94.27009}. The experiments on these materials show many surprising results which apparently differentiate them from other FeSC families (iron chalcogenide and iron pnictide). First, angle-resolved photoemission
(ARPES) studies indicate that there are only electron-like Fermi surfaces (FS) and no hole-like FS
in A$_{x}$Fe$_{2-y}$Se$_{2}$ \cite{NM.20.273,PRL.106.187001,PRL.106.107001}, while other FeSC families have both hole- and electron-like FS simultaneously \cite{AP.59.803}. Second, they are the first FeSC that approximates an antiferromagnetic (AFM) insulator accompanied by Fe-vacancy order \cite{EPL.94.27009}. And, the AFM order has a novel $\sqrt{5}\times\sqrt{5}$ block-type structure with an unprecedented high transition temperature of ${T}_{N}=559\mathrm{K}$ and a large magnetic moment of $3.31\mu_{B}/\mathrm{Fe}$ \cite{CPL.28.086104}. The implications of these anomalous features on superconductivity in A$_{x}$Fe$_{2-y}$Se$_{2}$ are not clear presently. And moreover, the questions about the pairing mechanism and the gap structure of the superconductivity in these systems still remain open.

From the theoretical viewpoint, the absence of hole Fermi pockets formally violates the nesting condition of the $s_{\pm}$ pairing symmetry, whose gap function changes signs between the hole and the electron pockets as in the most cases of other FeSC families, so that it requires a reconsideration of the itinerant spin-fluctuation mechanism in these new systems. Based on the multi-orbital Hubbard model and the functional-renormalization-group calculation, Wang \emph{et al.}\cite{EPL.93.57003} find that the $d_{x^{2}-y^{2}}$-wave instability is the leading pairing channel with a subleading extended $s$-wave component. A similar conclusion has also been obtained by Maier \emph{et al.}\cite{PRB.83.100515} through performing the random-phase-approximation (RPA) calculation. On the other hand, based on the local antiferromagnetic exchange interactions,  Seo \emph{et al.} \cite{PRX.1.011009} predict an $s$-wave pairing symmetry in this system that can account for the ARPES experimental results\cite{NM.20.273,PRL.106.187001,PRL.106.107001}. Khodas \emph{et al.} \cite{PRL.108.247003} suggest an $s^{+-}$ pairing symmetry, in which the gap changes sign between the hybridized pockets, by including the interpocket pairing.

Experimentally, the ARPES measurements have reported a nodeless superconducting (SC) gap on the large Fermi pockets around the zone corner in these materials \cite{NM.20.273,PRL.106.187001,PRL.106.107001,EPL.93.57001,PRB.83.140508}. In particular, the recent measurements further find an isotropic SC gap distribution on the small electron Fermi pocket around the Z point\cite{PRB.85.220504,EPL.99.67001}, which favors the $s$-wave pairing symmetry. The experimental evidences from the nuclear magnetic resonance (NMR) experiments suggest an $s^{+-}$-wave pairing symmetry \cite{PRB.83.104508,JPSJ.80.043708}, which is consistent with the isotropic gaps reported by ARPES. And, the muon spin spectroscopy experiment \cite{PRB.85.100501} suggest an $s$-wave symmetry and the results are incompatible with a clean $d$-wave model. In addition, the specific heat measurement also indicates that the gap is nodeless \cite{PRB.83.144511}. Though, the inelastic neutron scattering (INS) measurements  \cite{PRL.107.177005,PRB.86.024502} report a spin resonance in the superconducting state at $\bm{Q}\approx(\pi,\pi/2)$ which is very close to the value predicted by Maier \emph{et al.} \cite{PRB.83.100515} with a $d$-wave pairing symmetry, another spin excitation near $\bm{Q}=(0,\pi)$ was also observed in the superconducting state \cite{PRB.86.024502}. The latter is identical to that in many iron arsenide superconductors, which are believed to be the $s^{+-}$-wave superconductors.

In this paper, we will study the spin fluctuations and the paring symmetry in A$_{x}$Fe$_{2-y}$Se$_{2}$ with the fluctuation exchange (FLEX) approximation by including the next-nearest-neighbor (NNN) antiferromagnetic (AF) coupling $J_{2}$,
a localized magnetic interaction, into the multi-orbital Hubbard model which is treated within an itinerant electron picture. In fact, the INS experiments \cite{NP.5.555,PRB.84.054544,PRL.106.057004,NC.2.580} have indicated
that the NNN exchange coupling is AF and has a similar magnitude in many FeSC, although the nearest-neighbor (NN) exchange coupling is quite different. This implies that the NNN exchange coupling may play an important
role in the FeSC. But, in the FeSC with both hole- and electron-like FS the NNN exchange coupling has the similar effect with the FS nesting on the spin excitations. However, we will show that both the itinerant and the
localized natures of electrons are essential to describe properly the physical properties in A$_{x}$Fe$_{2-y}$Se$_{2}$, where the hole-like Fermi pocket disappears. In particular, a small $J_{2}$ can change the pairing gap from the $d$-wave symmetry to the $s$-wave symmetry, which is consistent with nodeless gap observed in recent experiments.  Considering the degeneracy of the two pairing states near the transition, we suggest a possible $s+id$-wave state in order to account for the spin resonance observed in INS experiments. We have also studied the real-space structures of the pairing gap at different orbits, from which we obtain the expressions for the favorable pairing gap functions.

\section{MODEL AND FLEX METHOD}

The model Hamiltonian consists of two parts:
\begin{equation}
H=H_{0}+H_{int}.
\end{equation}
The non-interacting part $H_{0}$ is described by the two-dimensional five-orbital tight-binding model in the unfolded (1-Fe) Brillouin zone (BZ) as introduced in Ref.\onlinecite{EPL.93.57003}. The parameters of the model involves the nearest- and second-neighbor
hopping among the Fe $3d$ orbits. The resulting tight-binding Hamiltonian reads
\begin{equation}
H_{0}=\sum_{\bm{k},\sigma}\sum^{5}_{\alpha,\beta=1}c^{\dag}_{\bm{k}\alpha\sigma}K_{\alpha\beta}(\bm{k})
c_{\bm{k}\beta\sigma},
\end{equation}
where $c^{\dag}_{\bm{k}\alpha\sigma}$ ($c_{\bm{k}\alpha\sigma}$) creates (annihilates) an electron with spin $\sigma$
and momentum $\bm{k}$ in the orbital $\alpha$. The parameters used in constructing $K_{\alpha\beta}(\bm{k})$ can be found in Ref.\onlinecite{EPL.93.57003}. However, in this paper, we use a gauge transformation on the five orbitals to change the orbital symmetry from ($d_{Z^{2}}$,$d_{XZ}$,$d_{YZ}$,$d_{X^{2}-Y^{2}}$,$d_{XY}$) to ($d_{z^{2}}$,$d_{xz}$,$d_{yz}$,$d_{xy}$,$d_{x^{2}-y^{2}}$), where $x$, $y$ refer to the nearest-neighbor Fe-Fe directions and
$X$, $Y$ refer to the diagonal directions. The transformation matrix is given by
\begin{eqnarray}
G=\left(
    \begin{array}{ccccc}
      1 & 0 & 0 & 0 & 0 \\
      0 & \frac{1}{\sqrt{2}} & -\frac{1}{\sqrt{2}} & 0 & 0 \\
      0 & \frac{1}{\sqrt{2}} & \frac{1}{\sqrt{2}} & 0 & 0 \\
      0 & 0 & 0 & 1 & 0 \\
      0 & 0 & 0 & 0 & -1 \\
    \end{array}
  \right),
\end{eqnarray}
and $K_{\alpha\beta}(\bm{k})$ can be obtained from $H_{\alpha\beta}(\bm{k})$ in Ref.\onlinecite{EPL.93.57003} through
the following transformation:
\begin{equation}
K(\bm{k})=G^{\dag}H(\bm{k})G.
\end{equation}
\begin{figure}
  \centering
  \includegraphics[scale=0.60]{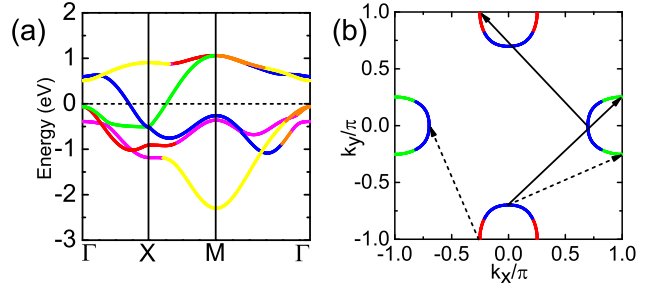}
  \caption{\label{fig1}(Color online) Energy band structure of the five orbital model (a) and the FS in the unfolded BZ for $0.3$ electron doping (b). The colors indicate the majority orbital character (red: $d_{xz}$, green: $d_{yz}$, blue: $d_{xy}$, yellow: $d_{x^{2}-y^{2}}$, pink: $d_{z^{2}}$, orange: $d_{xz}$ and $d_{yz}$ with the same weight). The arrows with solid lines in (b) indicate the weak nesting vectors $\bm{q}=(\pi,\pi)$ and $(-\pi,-\pi)$, the arrows with dashed lines indicate the weak nesting vectors $\bm{q}=(\pi,0.6\pi)$ and $(-0.6\pi,\pi)$.}
\end{figure}
The band structure and the associated FS for $0.3$ electron doping (6 electrons per Fe for parent compound) are shown in Fig. \ref{fig1}. We find that there are only electron FS around $(0,\pm\pi)$ and $(\pm\pi,0)$, which is consistent with the results of ARPES experiments \cite{NM.20.273,PRL.106.187001,PRL.106.107001}. Here, we do not consider the influence of Fe vacancies, because many experiments suggest that the antiferromagnetism with Fe vacancies and the superconductivity without Fe vacancies are in fact spatially separated \cite{NP.8.126,PRX.1.021020,PRB.84.180508,PRB.84.094504,PRB.85.100504,PRB.84.220505}.

The interactions between electrons are included in $H_{int}$ as
following,
\begin{align}
H_{int}&=U\sum_{i,\alpha}n_{i\alpha\uparrow}n_{i\alpha\downarrow}+V\sum_{i,\alpha\neq\beta,\sigma\sigma^{\prime}}
n_{i\alpha\sigma}n_{i\beta\sigma^{\prime}} \nonumber \\
&-J_{H}\sum_{i,\alpha\neq\beta}\bm{S}_{i\alpha}\cdot\bm{S}_{i\beta}
+J_{2}\sum_{\ll ij\gg,\alpha\beta}\bm{S}_{i\alpha}\cdot\bm{S}_{j\beta},
\end{align}
where $U$ ($V$) is the intra-orbital(inter-orbital) Coulomb interaction, $J_{H}$ the Hund's coupling, and $J_{2}$  the NNN antiferromagnetic coupling. $n_{i\alpha\sigma}=c^{\dag}_{i\alpha\sigma}c_{i\alpha\sigma}$ and
$\bm{S}_{i\alpha}=\sum_{\sigma\sigma^{\prime}}c^{\dag}_{i\alpha\sigma}\bm{\tau}_{\sigma\sigma^{\prime}}c_{i\alpha\sigma^{\prime}}$ are the density and spin operators respectively, where $\bm{\tau}$ are the Pauli matrices.

As mentioned above, due to the absence of the hole pockets in A$_{x}$Fe$_{2-y}$Se$_{2}$, the NNN antiferromagnetic coupling $J_{2}$ may play an essential role in determining the physical properties in these materials. Moreover, since the $J_{2}$ exchange coupling originates mostly from superexchange processes through Se, the effect of $J_{2}$ is underestimated  in the itinerant-electron model based on the pure iron lattice with only the on-site interactions. On the other hand, the INS experiments \cite{NP.5.555,PRB.84.054544,PRL.106.057004,NC.2.580} have indicated that the NNN exchange couplings are AF and have the similar values in many FeSC. Hence, we consider the effect of an admixture of both the on-site interactions and the $J_{2}$ exchange coupling here. A study based on the similar consideration for the interactions in the two-orbital model for pnictide superconductors with both electron and hole FS has been carried out from the strong coupling expansion \cite{PRL.106.217002}.

We carry out the investigation using the FLEX approximation\cite{Ann.Phys.93.206,*PRB.43.8044,NJP.11.025009,PRB.79.064517}, in which the Green's function and the spin/charge fluctuations are determined self-consistently. Here, we consider the following spin fluctuations $\hat{\chi}^{s}(\bm{q},i\omega_{n})$ and charge fluctuations $\hat{\chi}^{c}(\bm{q},i\omega_{n})$:
\begin{equation}
\hat{\chi}^{s}_{\alpha\beta}(\bm{q},i\omega_{n})=\int^{\frac{1}{T}}_{0}d\tau e^{i\omega_{n}\tau}\langle T_{\tau}[\bm{S}_{\alpha}(\bm{k},\tau)\cdot\bm{S}_{\beta}(-\bm{k},0)]\rangle,
\label{sf}
\end{equation}
\begin{equation}
\hat{\chi}^{c}_{\alpha\beta}(\bm{q},i\omega_{n})=\int^{\frac{1}{T}}_{0}d\tau e^{i\omega_{n}\tau}\langle T_{\tau}[n_{\alpha}(\bm{k},\tau)n_{\beta}(-\bm{k},0)]\rangle,
\label{cf}
\end{equation}
where $\bm{S}(\bm{k},\tau)=e^{\tau H}\bm{S}(\bm{k})e^{-\tau H}$ with $\bm{S}(\bm{k})$ the Fourier transformation of $\bm{S}_{i}$, $T$ is the temperature, $\tau$ and $\omega_{n}$ are respectively the imaginary time and  the Matsubara frequency. For the five-orbital model, the Green's function $\hat{G}$ and the self-energy $\hat{\Sigma}$ are expressed in a $5\times5$-matrix
form. As we use the spin and charge fluctuations expressed in Eq. (\ref{sf}) and Eq. (\ref{cf}), the susceptibility $\hat{\chi}^{0}$ and the effective interaction $\hat{V}$ also have a $5\times5$-matrix form. The Green's
function satisfies the Dyson equation
$\hat{G}(k)^{-1}=\hat{G}^{0}(k)^{-1}-\hat{\Sigma}(k)$, where the
self-energy is given by
$\Sigma_{mn}(k)=\frac{T}{N}\sum_{q}\sum_{\mu\nu}V_{n\mu,
m\nu}(q)G_{\mu\nu}(k-q)$ and the bare Green's function reads
$\hat{G}^{0}(k)=[\mathrm{i}\omega_{n}-K(\bm{k})+\mu]^{-1}$. Here, $k$ includes both momentum and frequency with
$k\equiv(\bm{k},i\omega_{n})$. The fluctuation exchange interaction is given by:
\begin{align}
\hat{V}(q)&=\frac{3}{2}\hat{U}_{s}(q)[(\hat{I}-\hat{\chi}^{0}(q)\hat{U}_{s}(q))^{-1}-\hat{I}]\hat{U}_{s}(q) \nonumber \\
&+\frac{1}{2}\hat{U}_{c}(q)[(\hat{I}+\hat{\chi}^{0}(q)\hat{U}_{c}(q))^{-1}-\hat{I}]\hat{U}_{c}(q) \nonumber \\
&+\frac{1}{2}(\hat{U}_{s}(q)\chi^{0}(q)\hat{U}_{s}(q)+\hat{U}_{c}(q)\chi^{0}(q)\hat{U}_{c}(q)),
\label{ev}
\end{align}
with the identity matrix $\hat{I}$ and the irreducible susceptibility $\chi^{0}_{\alpha\beta}(q)=-\frac{T}{N}\sum_{k}G_{\beta\alpha}(k+q)G_{\alpha\beta}(k)$.
The interaction
matrix for the spin(charge) fluctuation
$\hat{U}^{s}$($\hat{U}^{c}$) is given by: For $\alpha=\beta$, $U^{s}_{\alpha\beta}=U-4J_{2}\cos k_{x}\cos k_{y}$
($U^{c}_{\alpha\beta}=U$); For $\alpha\neq\beta$, $U^{s}_{\alpha\beta}=J_{H}-4J_{2}\cos k_{x}\cos k_{y}$
($U^{c}_{\alpha\beta}=4V$).

The Dyson equation, the self-energy and the interaction matrix Eq. (\ref{ev}) form a closed set of equations and can be solved self-consistently. Since the equations must be solved at fixed electronic density $n$, they are subject to the additional constraint:
\begin{equation}
n(T)=\frac{T}{N}\sum_{k}e^{i\omega_{n}0^{+}}\mathrm{Tr}\hat{G}(k,\mu(T))=n,
\end{equation}
where the chemical potential $\mu$ must be adjusted accordingly during the self-consistent calculation.

After obtaining the renormalized Green's function $\hat{G}$, we can
solve the "Eliashberg" equation,
\begin{eqnarray}
\lambda\phi_{\alpha\beta}(k)&=&-\frac{T}{N}\sum_{q}\sum_{\mu\nu}
V^{s/t}_{\alpha\beta}(q)G_{\alpha\mu}(k-q) \nonumber\\
& &\times G_{\beta\nu}(q-k)\phi_{\mu\nu}(k-q),
\label{eliashberg}
\end{eqnarray}
where the spin-singlet and spin-triplet pairing interactions
$\hat{V}^{s}$ and $\hat{V}^{t}$ are given by,
\begin{equation}
\hat{V}^{s}(q)=\frac{3}{2}\hat{U}^{s}\hat{\chi}^{s}(q)\hat{U}^{s}-
\frac{1}{2}\hat{U}^{c}\hat{\chi}^{c}(q)\hat{U}^{c}+\frac{1}{2}
(\hat{U}^{c}+\hat{U}^{s}),
\label{vs}
\end{equation}
\begin{equation}
\hat{V}^{t}(q)=-\frac{1}{2}\hat{U}^{s}\hat{\chi}^{s}(q)\hat{U}^{s}-
\frac{1}{2}\hat{U}^{c}\hat{\chi}^{c}(q)\hat{U}^{c}+\frac{1}{2}
(\hat{U}^{c}-\hat{U}^{s}).
\end{equation}
The most favorable SC pairing symmetry corresponds to the
eigenvector $\phi_{\alpha\beta}(k)$ with the largest eigenvalue $\lambda$.

Our numerical calculations are carried out on the $64\times64$ $\mathbf{k}$ meshes with 4096
Matsubara frequencies. In the calculations, we set the intra-orbital Coulomb interaction $U=1.2\mathrm{eV}$, the inter-orbital Coulomb interaction $V=0.35\mathrm{eV}$ and the Hund's coupling $J_{H}=0.2\mathrm{eV}$. The electron density is set as $6.3$ electrons per Fe and the temperature is $0.008\mathrm{eV}$.

\section{RESULTS AND DISCUSSION}

\subsection{Spin fluctuation}

\begin{figure}
  \centering
  \includegraphics[scale=0.60]{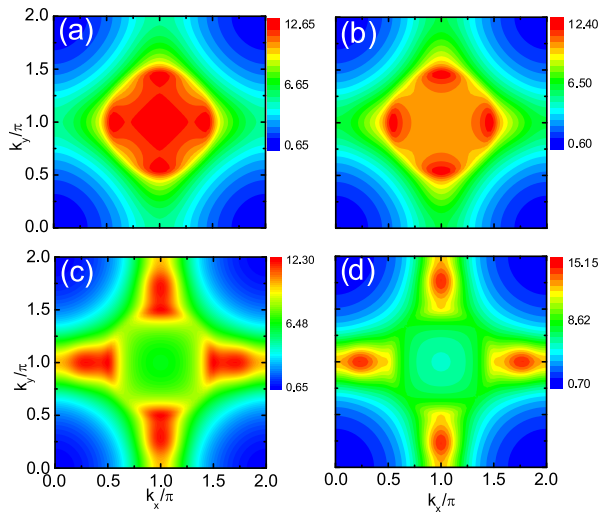}
  \caption{\label{fig2}(Color online) Physical spin susceptibilities for different NNN antiferromagnetic coupling $J_{2}$. (a) $J_{2}=0$, (b) $J_{2}=0.01 \mathrm{eV}$, (c) $J_{2}=0.04 \mathrm{eV}$ and (d) $J_{2}=0.05 \mathrm{eV}$.}
\end{figure}

As the spin fluctuations dominate over the charge fluctuations in our model described above, we only discuss the spin fluctuations and their contributions to the pairing interactions of electrons in this paper. We first study the evolution of the spin fluctuations with the NNN antiferromagnetic coupling $J_{2}$. In Fig. \ref{fig2}, we present the physical spin susceptibilities $\chi^{s}(\bm{q},\omega=0)=\sum_{\alpha\beta}\chi^{s}_{\alpha\beta}(\bm{q},\omega=0)$ for four different values of $J_{2}$. For $J_{2}=0$, the absence of the hole pocket around $\Gamma$ point in the unfolded BZ reomves the dominant $\bm{q}=(0,\pm\pi)$ and $(\pm\pi,0)$ nestings in iron pnictide, so that the peaks of the spin fluctuations at $(0,\pm\pi)$ and $(\pm\pi,0)$ disappear in A$_{x}$Fe$_{2-y}$Se$_{2}$. Instead, there are weak nestings at $\bm{q}=(\pm\pi,\pm0.6\pi)$, $(\pm0.6\pi,\pm\pi)$ and $(\pm\pi,\pm\pi)$ in this system (see Fig. \ref{fig1}(b)), which lead to the broad peak around $(\pi,\pi)$, together with the four satellite peaks at $(\pi,\pm0.6\pi)$ and $(\pm0.6\pi,\pi)$ (Fig. \ref{fig2}(a)). We note that these results are not consistent with the experimental results from INS measurements \cite{PRL.107.177005,PRB.86.024502}, in which the peak around $(\pi,\pi)$ are not observed. Thus, a simple Fermi surface nesting scenario within the itinerant-electron model is not enough to explain the neutron scattering experiments.
When turning on the $J_{2}$ interaction term, we find that the broad peak around $(\pi,\pi)$ decreases rapidly. Meanwhile, the intensity of the four satellite peaks increases and their positions move towards $(0,\pm\pi)$ and $(\pm\pi,0)$, as shown in Fig. \ref{fig2} (b)-(d) for $J_{2}=0.01\mathrm{eV}$, $0.04\mathrm{eV}$ and $0.05\mathrm{eV}$ respectively. As the $J_{2}$ term favors the diagonal antiferromagnetic order in the square lattice of Fe, it competes with the spin fluctuations coming from the nesting of FS with the nesting wavevector $(\pi,\pi)$ which is a kind of axial antiferromagnetic fluctuations. Thus, an enough large $J_{2}$ can suppress the spin fluctuations from the nesting effect of the FS. It is worthy to pointing out that we do not need a large value of $J_{2}$ comparing to the band width ($4\mathrm{eV}$) to change the character of the spin fluctuations. We find that when $J_{2}>0.04\mathrm{eV}$, the peak at $(\pi,\pi)$ is completely suppressed which is consistent with the results of INS experiments \cite{PRL.107.177005,PRB.86.024502}. And, the $(\pi,0)$ spin excitations observed in $\mathrm{Rb}_{0.82}\mathrm{Fe}_{1.68}\mathrm{Se}_{2}$ \cite{PRB.86.024502} also emerge for $J_{2}>0.04\mathrm{eV}$. In addition, for $0.03\mathrm{eV}<J_{2}<0.05\mathrm{eV}$, the peaks around $(\pm\pi,\pm0.6\pi)$ and $(\pm0.6\pi,\pm\pi)$ move to about $(\pm\pi,\pm0.5\pi)$ and $(\pm0.5\pi,\pm\pi)$, which is also consistent with the INS results. If we further increase $J_{2}$ to $J_{2}>0.05\mathrm{eV}$, the peaks around $(\pm\pi,\pm0.5\pi)$ and $(\pm 0.5\pi,\pm\pi)$ which comes mainly from the FS nesting disappear completely, and the four peaks of the spin fluctuations will now at the positions between $(\pi,0.5\pi)$ and $(\pi,0)$, and its symmetric points, so that it breaks down the consistence with the INS experiments. These results imply that an admixture of both the on-site interactions and the $J_{2}$ exchange coupling is appropriate and the $J_{2}$ exchange coupling has significant effects on the physical properties of A$_{x}$Fe$_{2-y}$Se$_{2}$. So, we will further study the pairing symmetries mediated by the spin fluctuations with the including of the $J_{2}$ term.

\subsection{Symmetry of the superconducting pairing}

\begin{figure}
  \centering
  \includegraphics[scale=0.36]{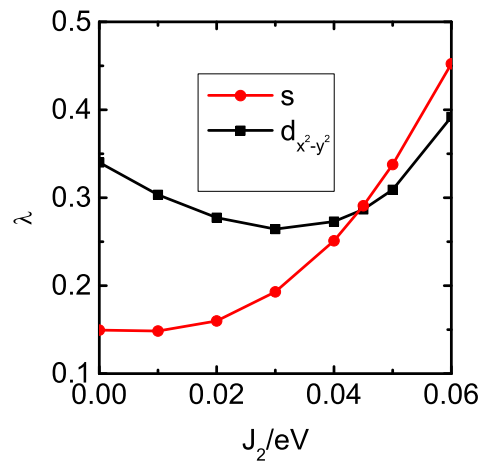}
  \caption{\label{fig3}(Color online) $J_{2}$ dependence of the eigenvalues for the $s$-wave and the $d$-wave solutions of the Eliashberg equation at $U=1.2\mathrm{eV}$, $V=0.35\mathrm{eV}$, $J_{H}=0.2\mathrm{eV}$ and $T=0.008\mathrm{eV}$.}
\end{figure}

 With the spin fluctuations, we can search for the most favorable pairing symmetry mediated by the spin fluctuations by solving the Eliashberg equation Eq.(\ref{eliashberg}). We find that the eigenvalue $\lambda$ has the maximum values in the spin-singlet channel and it is nearly zero in the spin-triplet channel in the range of parameters we have used. The paring functions corresponding to the two leading eigenvalues in the spin-singlet channel have the $s$-wave and $d$-wave symmetry. In Fig. \ref{fig3}, we show these eigenvalues as a function of $J_{2}$ for 0.3 electron doping corresponding to $\mathrm{K}_{0.6}\mathrm{Fe}_{2}\mathrm{Se}_{2}$. For a small $J_{2}$ ($<0.045\mathrm{eV}$), the eigenvalue for the $d$ wave is larger than that for the $s$-wave, so the $d$-wave pairing symmetry is dominant over the $s$-wave pairing symmetry. However, the eigenvalue for the $s$ wave
 pairing follows a monotonic increase with  $J_{2}$, and when $J_{2}>0.045\mathrm{eV}$ the $s$-wave pairing surpasses the $d$-wave pairing and becomes the most favorable pairing state. We note that the threshold value for $J_{2}$ is much smaller than that needed in the local-antiferromagnetic-exchange model \cite{PRX.1.011009} and is comparable to the experimental value from the INS measurement \cite{NC.2.580}.

\begin{figure}
  \centering
  \includegraphics[scale=0.60]{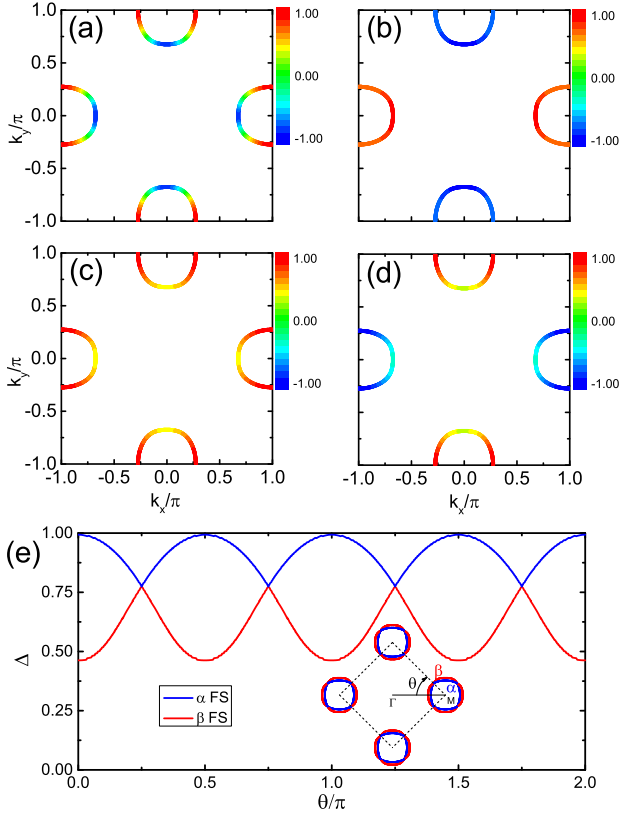}
  \caption{\label{fig4}(Color online) Pairing functions on FS. (a) and (b) are the $s$-wave and the $d$-wave solutions for $J_{2}=0$, respectively; (c) and (d) are the $s$-wave and the $d$-wave solutions for $J_{2}=0.05\mathrm{eV}$.  (e) is the variation of the gap magnitude along the FS for the $s$-wave pairing state shown in figure (c). The inset in (e) is the illustration of the FS in the folded BZ (dashed lines). The angle $\theta$ represents the position on the FS with respect to the $\Gamma-M$ direction. $\alpha$ labels the inner FS (blue), while $\beta$ labels the outer one (red). }
\end{figure}
\begin{figure}
  \centering
  \includegraphics[scale=0.60]{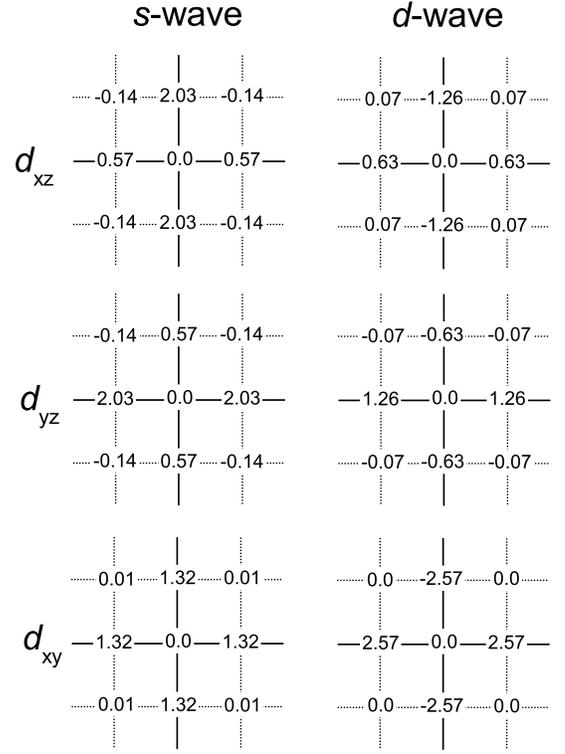}
  \caption{\label{fig5} Real-space structures of the pairing gaps at different orbits for the $s$-wave and the $d$-wave  symmetry for $J_{2}=0$. The values are shown with the relative intensities of the pairing gap. $d_{xz}$, $d_{yz}$ and $d_{xy}$ indicate the $t_{2g}$ orbits of Fe, which contribute mainly to the electronic states near the Fermi surface.}
\end{figure}
The gap functions on the FS for the $s$-wave and the $d$-wave symmetry corresponding to two representative values of $J_{2}=0$ and $0.05\mathrm{eV}$ are shown in Fig. \ref{fig4}. Here, the magnitudes of the gap have been normalized into the range from $-1$ to $+1$. In the case of $J_{2}=0$, the $s$-wave gap function has nodes (zero points of gap) on the FS, yet the dominant pairing symmetry here is the $d$-wave.  The gap of the $d$ wave symmetry for $J_{2}=0$ [Fig. \ref{fig4} (b)] has no node on the FS, this is consistent with the ARPES results of a nodeless SC gap on the large Fermi pockets around the zone corner \cite{NM.20.273,PRL.106.187001,PRL.106.107001,EPL.93.57001,PRB.83.140508}. But, it has line nodes along the diagonal direction in the Brillouin zone. Recent ARPES experiments further report an isotropic SC gap on the small electron Fermi pocket around the Z point\cite{PRB.85.220504,EPL.99.67001}, so it rules out the $d$-wave symmetry. In addition, as argued by Mazin \cite{PRB.84.024529}, this $d$-wave gap is rather fragile when the hybridization between the electron pockets in the folded BZ is included. For the same reason, the $d$-wave gap for $J_{2}=0.05$ [Fig. \ref{fig4} (d)] is also fragile.  The only pairing symmetry that is stable and consistent with experiments is the $s$-wave gap in the case of $J_{2}=0.05$ [Fig. \ref{fig4} (c)], which is also the dominant pairing symmetry for $J_{2}=0.05$ as shown in Fig. \ref{fig3}. We note that the gap magnitude of this $s$-wave state has in fact a variation along the FS, though it has no node. To compare with the ARPES experiments, in Fig. \ref{fig4} (e), we present the variation of the gap along the FS in the folded (2-Fe) BZ. We find that the gap on the inner FS is larger than that on the outer FS and it has a relative small variation (about 25\%) along the FS.

Though the nodeless $s$-wave gap is consistent with the ARPES experiments, it can not account for the spin resonance observed in the neutron scattering\cite{PRL.107.177005,PRB.86.024502}, because a sign change between the gaps connecting by the wavevector at which the resonance occurs is essential to a nonzero spin response\cite{PRB.53.5149,PRB.58.2895}. However, we note that the $s$-wave and $d$-wave states are nearly degenerate around $J_{2}=0.045\mathrm{eV}$, as seen from Fig. \ref{fig3}. In this case, an $s+id$ pairing state may be expected, which has been discussed in the early days of Fe pnictides \cite{PRL.102.217002}. It has been shown that the mixed state with $s+id$ pairing symmetry does give rise to the spin resonance observed in the INS experiments \cite{arxiv.1206.5235}. On the other hand, the gap magnitude for the $s+id$ pairing function is $\Delta(\bm{k})=\sqrt{\Delta^{2}_{s}(\bm{k})+\Delta^{2}_{d}(\bm{k})}$, where $\Delta_{s}(\bm{k})$ and $\Delta_{d}(\bm{k})$ are the gaps for the $s$-wave and the $d$-wave states, respectively. It is nodeless, consistent with the ARPES results \cite{NM.20.273,PRL.106.187001,PRL.106.107001,EPL.93.57001,PRB.83.140508}.

Because the spin fluctuation is dominant over the charge fluctuation, the pairing interaction in the spin-singlet channel is positive (repulsive) [see Eq. (\ref{vs})]  and has a maximum around the wave vectors $\bm{Q}$ at which the spin fluctuation has a peak. For a repulsive pairing interaction, the most favorable SC gap must satisfy the condition $\Delta(\bm{k})\Delta(\bm{k}+\bm{Q})<0$ in order to get the largest eigenvalue $\lambda$ of the "Eliashberg" equation, as can be seen from Eq. (\ref{eliashberg}). In the case of  $J_{2}=0$, the pairing interaction comes mainly from the spin fluctuations resulting from the nesting of the FS. As demonstrated above in Fig. \ref{fig2} (a), two sets of spin fluctuations have been observed, with peaks around $\bm{q}=(\pm\pi,\pm0.6\pi)$, $(\pm0.6\pi,\pm\pi)$ and $(\pm\pi,\pm\pi)$ corresponding to the nesting wavevectors shown in Fig.\ref{fig1}(b). Therefore, the favorable gaps on FS connected by these nesting vectors will change the signs. This gives rise to two possibilities, namely the $s$-wave and the $d$-wave symmetry, as shown in Fig. \ref{fig4} (a) and (b). Energetically, the $d$-wave state here will be more favored because it opens a full gap around the Fermi pockets[Fig. \ref{fig4}(b)]. Physically, the $J_{2}$ term will induce the electron pairings along the NNN bonds, which gives rise to the pairing function of the form $\cos k_{x}\cos k_{y}$, i.e., the so-called $s^{+-}$ pairing in the case of both electron and hole pockets. In the absence of the hole pockets here, it exhibits itself as the $s$-wave gap function as shown in Fig.\ref{fig4}(c).

\begin{figure}
  \centering
  \includegraphics[scale=0.60]{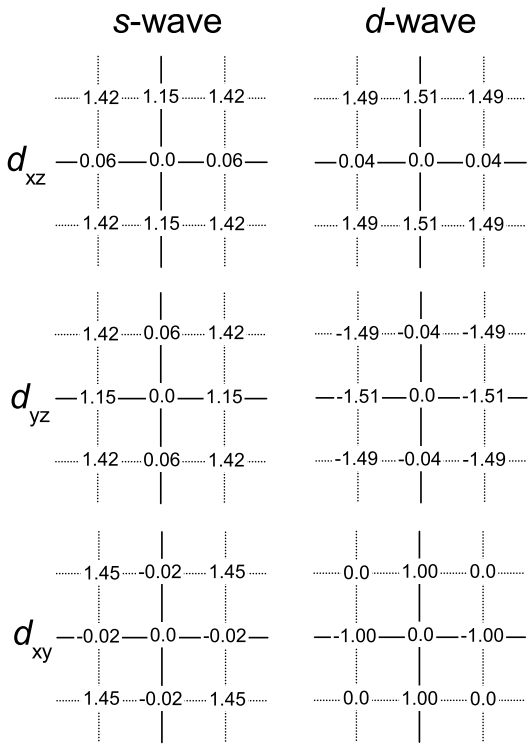}
  \caption{\label{fig6} Real-space structures of the pairing gaps at different orbits for the $s$-wave and the $d$-wave symmetry for $J_{2}=0.05\mathrm{eV}$. Notions are the same as in Fig. \ref{fig5}.}
\end{figure}

In order to quantify the detail gap function and gain more understanding of the physical picture discussed above, we will consider the following Fourier transformation of the gap function:
\begin{equation}
\Delta_{\alpha\beta}(\bm{R})=\frac{1}{N}\sum_{\bm{q}}e^{i\bm{q}\cdot\bm{R}}\Delta_{\alpha\beta}(\bm{q})
\end{equation}
 at the lowest Matsubara frequency $\omega_{n}=\pi T$, where $\bm{R}$ is the real-space vector, $\alpha$ and $\beta$ are the orbital indices. We find that the inter-orbital ($\alpha\neq\beta$) pairing magnitudes are negligible in comparison with the intra-orbital ($\alpha=\beta$) pairings, suggesting that the superconducting gaps come mainly from the intra-orbital pairings. In addition, the pairing magnitudes in the $e_{g}$ ($d_{z^{2}}$ and $d_{x^{2}-y^{2}}$) orbits are also much smaller than those in the $t_{2g}$ ($d_{xz}$, $d_{yz}$ and $d_{xy}$) orbits, because the contribution to the FS are mainly from the $d_{xz}$, $d_{yz}$ and $d_{xy}$ orbits. Thus, we will only consider the gap properties in the $t_{2g}$ orbits in the following.

 The real-space structures of the gap functions $\Delta_{\alpha\beta}(\bm{R})$ in the $t_{2g}$ orbits up to the next nearest neighbors for $J_{2}=0$ and $J_{2}=0.05\mathrm{eV}$ are plotted in Fig. \ref{fig5} and Fig. \ref{fig6}, respectively. The general features are: i) the on-site paring magnitude is always zero, simply because the strong on-site Coulomb repulsion prohibits the on-site pairing. ii) The magnitudes of the $d_{xz}$ and the $d_{yz}$ orbits have only the $C_{2}$ symmetry due to the orbital character. In the case of $J_{2}=0$, the largest magnitude of the gap function appears at the $d_{xy}$ orbit in the $d$-wave pairing channel. It is nearly twice as those at the $d_{xz}$ and $d_{yz}$ orbits in the same channel, and is also larger than those in the $s$-wave paring channel, so this pairing state is dominant. In this case, the pairing at the $d_{xy}$ orbit occurs only along the NN bonds and the pairing amplitude changes sign between the $x$ and $y$ directions, so it gives rise to the usual $d$-wave gap function of the form $\cos k_{x}-\cos k_{y}$. The induction of the pairings along the NNN bonds due to the $J_{2}$ term is obvious from Fig. \ref{fig6}. But, the important feature is that, only at the $d_{xy}$ orbit in the $d$-wave channel, is the NNN bond pairing not induced at all, all others are nearly induced equally and these pairing are larger than those along the NN bonds. Thus, the $s$-wave pairing is favored in the case of $J_{2}=0.05\mathrm{eV}$, which has the gap function of the form $\cos k_{x}\cos k_{y}$ due to the parings along the NNN bonds.

From the real-space distributions of the pairing bonds, we can obtain the expressions of the pairing gaps at different orbits. For $J_{2}=0$, the favorable pairing has the $d$-wave symmetry and the pairing functions are,
\begin{align}
\Delta_{xz}(\bm{k})&=\Delta_{0}[0.63\cos(k_{x})-1.26\cos(k_{y}) \nonumber \\
&+0.14\cos(k_{x})\cos(k_{y})],
\end{align}
\begin{align}
\Delta_{yz}(\bm{k})&=\Delta_{0}[1.26\cos(k_{x})-0.63\cos(k_{y}) \nonumber \\
&-0.14\cos(k_{x})\cos(k_{y})],
\end{align}
\begin{equation}
\Delta_{xy}(\bm{k})=\Delta_{0}[2.57\cos(k_{x})-2.57\cos(k_{y})],
\end{equation}
 While, for $J_{2}=0.05\mathrm{eV}$, the favorable pairing has the $s$-wave symmetry and the pairing functions are,
\begin{align}
\Delta_{xz}(\bm{k})&=\Delta_{0}[0.06\cos(k_{x})+1.15\cos(k_{y}) \nonumber \\
&+2.84\cos(k_{x})\cos(k_{y})],
\end{align}
\begin{align}
\Delta_{yz}(\bm{k})&=\Delta_{0}[1.15\cos(k_{x})+0.06\cos(k_{y}) \nonumber \\
&+2.84\cos(k_{x})\cos(k_{y})],
\end{align}
\begin{align}
\Delta_{xy}(\bm{k})&=\Delta_{0}[-0.02\cos(k_{x})-0.02\cos(k_{y}) \nonumber \\
&+2.90\cos(k_{x})\cos(k_{y})],
\end{align}
 These pairing functions may be helpful for the experimental analysis of the pairing states.

\section{Summary}

In summary, we have investigated the spin fluctuations and the pairing symmetry in A$_{x}$Fe$_{2-y}$Se$_{2}$ which has only electron Fermi pockets, by the fluctuation exchange approximation. Besides the on-site interactions, the next-nearest-neighbor antiferromagnetic coupling $J_{2}$ is also included. We find that both the itinerant and the localized natures of electrons are important to  describe the recent experimental results of the spin fluctuations and the pairing symmetry in these materials. In particular, a small $J_{2}$ can change the pairing gap from the $d$-wave symmetry to the $s$-wave symmetry, which is consistent with the nodeless gap observed in recent experiments. In addition, we have studied the real-space structures of the pairing gap at different orbits, from which we obtain the expressions for the favorable pairing gap functions.

\begin{acknowledgments}
This work was supported by the National Natural Science Foundation of China (91021001, 11190023 and 11204125)
and the Ministry of Science and Technology of China (973 Project Grants No.2011CB922101 and No. 2011CB605902).
\end{acknowledgments}

\bibliography{KFeSe}

\end{document}